\pdfoutput=1

\documentclass[12pt,preprint]{aastex}

\slugcomment{}

\shorttitle{Density Stratification and Rotation}
\shortauthors{Evonuk}

\begin{document}


\title{The Role of Density Stratification in Generating Zonal Flow Structures in a Rotating Fluid}


\author{Martha Evonuk}
\affil{Institut f\"{u}r Geophysik, ETH Z\"{u}rich, 8093 Z\"{u}rich, Switzerland}
\email{mevonuk@erdw.ethz.ch}



\begin{abstract}
Local generation of vorticity occurs in rotating density-stratified fluids as fluid parcels move radially, expanding or contracting with respect to the background density stratification. Thermal convection in rotating 2D equatorial simulations demonstrates this mechanism. The convergence of the vorticity into zonal flow structures as a function of radius depends on the shape of the density profile, with the prograde jet forming in the region of the disk where the greatest number of density scale heights occurs. The number of stable jets that form in the fluid increases with decreasing Ekman number and decreases with increasing thermal driving. This local form of vorticity generation via the density stratification is likely to be of great importance in bodies that are quickly rotating, highly turbulent, and have large density changes, such as Jovian planets. However, it is likely to be of lesser importance in the interiors of planets such as the Earth, which have smaller density stratifications and are less turbulent.
\end{abstract}





\section{Introduction}

Rotation plays a key role in fluid motions in both the liquid interiors of planets and planetary atmospheres. Much work has been done in the context of thin atmospheres and the Beta-effect \citep{williams79,williams85,williams03,cho96,allison00}. The effect of rotation and its role in producing zonal flow patterns as a function of latitude, such as seen on the giant gas planets, has been studied by several groups in the Boussinesq, constant background density, approximation \citep{christensen01,christensen02,aurnou01,aurnou04,heimpel05}; as well as in the anelastic, stratified background density, approximation \citep{glatzmaier05}. While studies in 3D and in the 2D meridional plane (Jones et al 2003; Rotvig and Jones 2006) reproduce jet structures similar to those seen in giant planets, these simulations are run in parameter regimes far from those present in planetary bodies. In particular, due to numerical constraints, these simulations are executed with viscosities and diffusivities that are too high, resulting in more diffusive and more laminar flows than should be present in the planets. While some of these simulations \citep{glatzmaier05,heimpel05} display some turbulent behavior they are still not as fully turbulent as we expect planets, such as Jupiter and Saturn, to be. This means that simulations in this more laminar regime interact with their boundaries more than they would in a more turbulent regime. Interactions with the boundaries lead to vorticity generation via the global mechanism of Busse rolls \citep{proudman16,busse83,busse94,busse02}, an effect that should become weaker as the fluid becomes more turbulent. Extrapolations have been made to project trends seen in simulations to higher Rayleigh numbers and lower Ekman numbers and thus into the parameter regime proper for planetary bodies \citep{christensen06}, however these extrapolations are made across many orders of magnitude and therefore may not hold. Further, the bulk of these studies have been conducted with constant background density while the planets exist in the density-stratified regime. As direct 3D simulations of fully turbulent bodies are not possible this paper seeks to expand the explored parameter regime in 2D and to look in particular at the effect of density stratification on the fluid behavior. These 2D simulations are in the equatorial plane without additional terms to add the effect of curvature of the boundaries in the northern and southern hemispheres. This essentially eliminates the effect of these boundaries and therefore any contribution to the vorticity from Busse roll structures. These simulations also span the full equatorial plane and do not include a core, eliminating the effect of a solid core on the fluid behavior. Therefore the vorticity and any jets formed will be purely the consequence of the local interaction of the rotation with the density stratification.

To review how jets form with latitude via interactions with the boundaries in 3D, I will first present vorticity generation via the geostrophic approximation. Next the effects of density will be discussed to show how the density stratification can provide a local source of vorticity. This local mechanism, in the context of the 2D equatorial plane, will be contrasted to the Boussinesq, constant density case. A suite of density stratified numerical simulations will be presented exploring a range of rotation rates (Ekman numbers) and driving rates (Rayleigh numbers) to explore the effect of rotation on the number of jets that are created, as a function of radius in the fluid. While the parameter regimes expected for planetary interiors cannot be reached even in these 2D simulations we will see that the results point towards the growing importance of this local mechanism of vorticity generation, especially in the regime appropriate for giant planets.




\section{Generating Vorticity}

\subsection{The Geostrophic Approximation} \label{geo}

Of foremost importance in this discussion is the momentum equation,
\begin{equation}
\frac{\partial}{\partial t}(\bar{\rho} {\bf u}) = -\nabla \cdot \left[ \bar{\rho} u_i u_j + P \delta_{ij} - 2 \nu \bar{\rho} \left(  e_{ij} - \frac{1}{3} (\nabla \cdot {\bf u}) \delta_{ij}    \right) \right] -\rho \bar{g} \hat{r} + 2 \bar{\rho} {\bf u} \times {\bf \Omega} ,
\end{equation}
written here in terms of the anelastic approximation with $t$ being time, $\rho$ the density, ${\bf u}$ the velocity vector, $P$ the pressure, $\delta_{ij}$ the Kronecker delta, $\nu$ the viscous diffusion, $e_{ij}$ the viscous stress tensor, $g$ the gravity, $\hat{r}$ the unit vector in the radial direction, and $\Omega$ the rotation rate. Quantities with an over bar indicate a background radial profile. If we assume that the pressure gradient and the Coriolis terms are the dominant terms in the momentum equation, and that the background density ($\rho_o$) is constant, then we get the following relation often referred to as the Geostrophic Approximation: $\nabla P = 2 \rho_o {\bf u} \times {\bf \Omega}$. Taking the curl of this equation gives the Taylor-Proudman theorem $\partial {\bf u} / \partial z = 0$, which dictates to first order that the change in the velocity parallel to the axis of rotation is zero. Therefore vertical columns spanning the convecting fluid are expected to form. These types of columns have been observed in laboratory experiments \citep{carrigan83,busse94} and have been implied to exist in the EarthÕs core from magnetic observations \citep{gubbins87}. How vorticity is generated in this situation can be viewed in the following way: as the column moves toward or away from the axis of rotation its height is forced to change due to the curvature of the bounds of the convective region. The columns change their width to conserve mass, as a column moves towards the axis of rotation it becomes taller and thinner, while as it moves away from the axis it becomes shorter and wider. This, coupled with the Coriolis force, generates positive vorticity in the case of the lengthening column and negative vorticity in the case of the shortening column. If we now add the sense of curvature of the boundary it can be shown \citep{busse02} that in the case of concave curvature, which applies to the outer boundaries of the planets, these columns will become tilted perpendicular to the axis of rotation, allowing the convergence of positive vorticity, prograde flow, on the equator-side of the columns, and negative vorticity, retrograde flow, on the side of the columns towards the axis. Pairs of these Busse rolls can then set up alternating prograde and retrograde flows at the surface as a function of latitude with the primary prograde flow at the equator similar to the flows observed in the surface clouds of the giant planets \citep{busse94}. This flow pattern is seen in Boussinesq simulations \citep{heimpel05} and arguably in anelastic simulations as well \citep{glatzmaier05}. However, as the level of turbulence in the fluid rises it becomes increasingly unlikely that columns spanning the convective zone can be maintained.  The Geostrophic Approximation will no longer be valid as the fluid becomes more turbulent and the inertial terms play a more important role. However, there is another mechanism that acts locally to generate vorticity that may gain increasing importance as this global mechanism breaks down in the high turbulence regime.

\subsection{The effect of density stratification} \label{den}

All convecting planetary interiors are density stratified to some extent. The density of the EarthÕs outer metallic core decreases on the order of 23\% from bottom to top \citep{boehler96}. This translates to 0.2 density scale heights where the number of density scale heights is defined as $N_{\rho} = \ln(\rho_b/\rho_t)$, with $\rho_b$ and $\rho_t$ being the densities at the bottom and top of the convective zone respectively. Other planets experience much greater changes in density through their fluid interiors. The density in JupiterÕs hydrogen and helium envelope changes by over 200 times if the upper level of the convective zone is taken to be at a pressure of 1000 bars, or $10^8$ Pa \citep{guillot99}. This translates to at least 5.3 density scale heights; naturally, as more of the upper atmosphere is included a greater number of density scale heights will occur. This change in the density does not go unnoticed by the fluid. Simple 2D simulations in a box show that a significant asymmetry in the fluid flow occurs in the density-stratified case, while in the constant density case symmetric flow is observed \citep{evonuk04}. As fluid plumes in the density-stratified case rise they move into less dense regions and expand, forming mushroom shaped plumes, while the downwelling plumes are compressed. Meanwhile, for constant density cases upwelling and downwelling plumes are in a sense indistinguishable. An additional asymmetry in the density-stratified case appears in the velocities. In denser regions the fluid moves more slowly than in less dense regions. When these asymmetries are coupled to the rotation some interesting effects become apparent (the following is a review of material to be covered in more detail in Glatzmaier et al.\ 2008, in preparation). In particular expansion and contraction velocities of rising and sinking fluid parcels act via the Coriolis force to produce vorticity. Rising expanding parcels generate negative vorticity while sinking contracting parcels generate positive vorticity. This mechanism is an entirely local process so there is no longer a need for coherent columns spanning the fluid. Instead, many small disconnected fluid columns can form aligned along the axis of rotation and their individual radial movements will generate vorticity. We can examine how this mechanism works in more detail in the limit of two dimensions in the equatorial plane. To look at the time rate of change of the vorticity due to rotation we take the curl of the Coriolis force, $\nabla \times (2 {\bf u} \times {\bf \Omega}) = - 2 {\bf \Omega} (\nabla \cdot {\bf u} )$. In the Boussinesq case where the density is constant there is no contribution from the Coriolis term to the vorticity as $\nabla \cdot {\bf u} =0$. However, in the anelastic case, which includes a density-stratification,
\begin{equation}
\nabla \cdot (\bar{\rho} {\bf u}) = 0, 
\end{equation}
we see that a non-zero term is produced by this cross product: 
\begin{equation}
\nabla \cdot {\bf u} =\frac{1}{\bar{\rho}}\nabla \cdot (\bar{\rho} {\bf u}) - \frac{1}{\bar{\rho}} \frac{d\bar{\rho}}{dr} u_r 
\end{equation}
so that
\begin{equation}
\nabla \times (2 {\bf u} \times {\bf \Omega}) = - 2 {\bf \Omega} (\nabla \cdot {\bf u} ) = 2 {\bf \Omega} \frac{1}{\bar{\rho}} \frac{d\bar{\rho}}{dr} u_r = 2 {\bf \Omega} \frac{d \ln \bar{\rho}}{dr} u_r.
\end{equation}
This term depends on the rate of rotation, the radial velocity, and the local inverse density scale height, $h_{\rho} = d \ln \bar{\rho} / dr$. For a stably stratified fluid where the density decreases with radius, the inverse density scale height is negative everywhere but changes in amplitude depending on the shape of the density curve. In a fluid where the density drops off more quickly towards the surface of the planet (Figure 1a) peak values of the inverse density scale height are seen near the surface of the planet. Conversely, if the density drops off more slowly near the surface than at depth, then peak values of the inverse density scale height occur at depth. This plays an important role in the convergence of the angular momentum. A simple schematic (Figure 1b, c) shows that fluid parcels leaving both the inner and outer regions in the equatorial plane both ultimately transfer negative vorticity away from the boundaries as the sinking contracting parcel and the rising expanding parcel both veer in the direction retrograde to that of rotation. If we consider a density profile where the peak negative values of the inverse density scale height occur at the upper boundary, or near the surface, then the transport of negative vorticity from the upper boundary will be dominant, resulting in a prograde flow near the surface of the disk and a retrograde flow at depth. Conversely, if the peak inverse density scale height occurs at depth then a retrograde flow will form near the surface.

While Busse rolls are unlikely to exist in the turbulent interiors of giant planets, in the more laminar anelastic simulations in 3D we expect to see both the global process of Busse rolls and the local process via the density stratification to play a role in the formation of zonal flow structures. Since these processes are difficult to separate in 3D it is of interest to look at the effect of the density change in 2D to eliminate Busse rolls and to approach more turbulent regimes.

\section{Numerics}

Thermal-convection simulations are performed using the Evonuk-Glatzmaier finite-volume code on a Cartesian grid in 2D \citep{evonuk06}. The basic equations solved are the momentum equation (Eq. 1), conservation of mass (Eq. 2), and the energy equation in terms of the entropy $S$ as shown below:
\begin{equation}
\frac{\partial}{\partial t}(\bar{\rho} S) = -\nabla \cdot \left( \bar{\rho}(S + \bar{S}) - \kappa_T\bar{\rho} \nabla S - \frac{C_P \kappa_R \bar{\rho}}{\bar{T}} \nabla T \right) + \frac{\bar{\rho}}{\bar{T}} \frac{d \bar{T}}{dr} \left( \kappa_T \frac{\partial S}{\partial r} + \frac{C_P \kappa_R}{\bar{T}} \frac{\partial T}{\partial r} \right)+\bar{\rho} Q_s.
\end{equation}
In this equation $\kappa_T$ is the thermal diffusivity, $C_P$ is the specific heat capacity at constant pressure, $\kappa_R$ is the radiative diffusivity, $T$ is the temperature, and $Q_s$ is the heating function. In these simulations the rotation rate, $\Omega$, and the rate of heating, $Q_s$, were varied to explore a range of parameters. The use of the finite volume method allows the inner core to be easily removed from the simulations to focus on the effect of rotation and the density stratification as there are no numerical issues associated with the origin point. Over four density scale heights were included along the profile shown in Figure 1a with numerical values appropriate for the interior of a Jovian planet with no core \citep{guillot99} but with a shape that is likely to also be appropriate for an Earth-like planet without a core. In the context of the above discussion this will imply a convergence of positive vorticity or prograde flow near the surface. To prevent the convecting fluid from splashing against the outer boundary of the simulation or feeling the jaggedness of the outer boundary, a stable buffer zone was added in the outer 10\% of the disk. Simulation results will be viewed looking down from the north so that eastward, or prograde motion, is in the counterclockwise direction. Simulations were performed with grid resolutions from $500\times500$ to $1600\times1600$.

There are three primary dimensionless numbers which can be used to characterize the equations. The Rayleigh number,
\begin{equation}
Ra = \frac{g\Delta S D^3}{C_P \nu \kappa_T},
\end{equation}
provides a measure of the buoyancy terms versus the diffusive terms. Here it is written in terms of the change in entropy, $\Delta S$, of the fluid in the convecting region. As there is no inner core, $D$ is the radius of the disk and the total change in entropy is calculated from the output of the simulation. Higher Rayleigh numbers are generally indicative of higher driving and more turbulent flows. The Ekman number, 
\begin{equation}
Ek = \frac{\nu}{2 \Omega D^2},
\end{equation}
provides a comparison of the viscous terms to the Coriolis term. Smaller Ekman numbers indicate fluids where the rotation is more dominant in determining fluid behavior. The Prandtl number, $Pr = \nu / \kappa_T$, is the a ratio of the viscous diffusivity to the thermal diffusivity. In all of the simulations the Prandtl number was taken to be $1.0$ and the diffusivities were prescribed to be constant. A fourth dimensionless number, the Reynolds number, $Re = U D / \nu$, compares the inertial terms to the viscous terms, providing a non-dimensional way of looking at the time-averaged mean Euclidean norm of the velocity, U, of the fluid. 

\section{Results}

The fluid flow patterns for three representative cases are shown in Figure 2. The middle panel shows a case with two jets, as hypothesized by the above discussion with an exterior prograde jet and an interior retrograde jet. A case with a smaller rate of rotation (also characteristic of a larger rate of driving) shows the flow moving directly through the interior forming a dipolar flow structure (Figure 2c). This flow pattern occurs when the effect of rotation is too small, with respect to the driving, to form jets. Conversely, at higher rotation rates or lower driving, a third jet can form (Figure 2a). The amplitudes of the jets in these representative cases are shown in Figure 3 which shows the time and longitudinally-averaged angular velocity (i.e. the tangential velocity divided by the radius) scaled to the rotation rate for the cases with three jets and two jets in Figure 2. These families, grouped by number of jets can be seen in Figure 4a showing the number of jets as a function of Ekman and Reynolds numbers. In agreement with Figure 2, as the rotation rate increases (Ekman number decreases) more jets are formed, whereas when the fluid is driven harder and the velocities increase (Reynolds number increases) fewer jets are formed.

It is also of interest to look the directionality of the velocities, that is, the ratio of the time-averaged maximum radial ($u_r$) to maximum tangential ($u_{\phi}$) velocities. This ratio is shown in Figure 4b plotted against the Ekman number. As the rotation rate increases, the vertical convection is suppressed and a greater fraction of the fluid flow is aligned in the tangential direction, as would be expected when more jets are formed. As large-scale vertical convection is suppressed, smaller scale vortices develop to transport energy radially, an effect shown in \citet{evonuk06}. This graph shows a very clear division between the number of jets with decreasing Ekman numbers and the ratio of maximum radial velocity to tangential velocity.

The convective heat flux, $F_{c} = C_P \bar{\rho} T u_r$ is a function of the temperature perturbation and the radial velocity, while the diffusive heat flux, $F_{d} = -C_P \bar{\rho} \kappa_T \partial (\bar{T} + T) / \partial r$ depends primarily on the total temperature, $\bar{T} + T$. The relationship between the Ekman number and the ratio of the mean convective heat flux to the mean diffusive heat flux is therefore very similar in distribution (though not in value) to that of the relationship between the Ekman number and the Reynolds number (Figure 4a), with larger ratios of convective heat flux to diffusive heat flux corresponding to higher Reynolds numbers. The Rayleigh number versus the ratio of the mean heat fluxes shows a very strong correlation between the two quantities as higher Rayleigh numbers mean larger driving and therefore larger radial velocities (Figure 4c). In this plot there appears to be no meaningful distinction based purely on number of jets present in the simulations.

Lastly, Figure 4d shows the Ekman number versus the Rayleigh number. Again, as the Ekman number decreases, more jets develop, while as the Rayleigh number increases (buoyancy increases due to increased driving), fewer jets form. Since planets are in a parameter regime of very low Ekman number and very high Rayleigh number it is of interest to see where they would lie on this plot. Figure 5 expands this graph to include representative values for the EarthÕs liquid outer core and Jupiter for comparison. It is obvious that even 2D simulations are far from the parameter regimes expected for these planets. Any trends chosen need to be extrapolated over many orders of magnitude. Two possible extrapolations have been included in Figure 6 to demonstrate this sensitivity. A conservative extrapolation places both the Earth and Jupiter in a regime of three jets, or a region where local vorticity generation is very important. However a trend could also be chosen to put them both in the transition area between the three-jet and two-jet regions, where this local mechanism becomes less important.

Also of importance when interpreting these results for individual planets is the extent of the density stratification. These simulations have been done with four density scale heights. Therefore we expect a graph appropriate for the Earth, with 0.2 density scale heights, to have the EarthÕs location on the graph shifted to the right with respect to the observed pattern. This would shift the Earth into the region of parameter space where the density-stratification plays a smaller role in generating vorticity. Conversely, Jupiter has a greater number of density scale heights; this will shift Jupiter to the left with respect to this pattern placing Jupiter more firmly in the parameter space where local vorticity generation, and the inclusion of the planetary density stratification, is very important.

\section{Conclusions}

This paper presents a local form of vorticity generation via interactions of the fluid moving through the background density profile with the Coriolis term. Simulations in 2D show this mechanism organizing the fluid into zonal flows with respect to radius in the equatorial plane with more jets forming as the rotation rate is increased. As the rotation rate is increased the radial convection of the fluid is suppressed and the energy moves radially via small convective cells imbedded in the zonal flow, as highlighted in \citet{evonuk06}. High driving suppresses the effect of rotation and results in a lower number of jets. This form of local vorticity generation is likely to be less important in planets that have small density stratifications and less turbulent flow, such as in the EarthÕs core. However in giant planets where there are a large number of density scale heights and the flow is very turbulent this form of vorticity generation is likely to play an important role. Therefore, in addition to asymmetries in the fluid flow structures and velocities, local vorticity generation provides another argument for including the density-stratification when modeling the fluid behaviors and magnetic field generation particularly in giant planets.

Lastly, it is important to keep in mind that in 3D, as well as in 2D, we are still far from the parameter regimes of planetary interiors and should therefore be cautious when extrapolating current modeling results to planetary interiors.

\acknowledgments

I would like to thank Jon Aurnou for several helpful discussions on the topic as well as Chris Finlay and Gary Glatzmaier. Thanks also to Philippe Cardin for inviting me to give a talk at IUGG and allowing me to attend the Les Houches Dynamo workshop, both of which helped me organize my thoughts on the subject. Computations were run on the UCSC Beowulf cluster, upsand, and on the CSCS cluster, palu, through ETH.




\clearpage



\begin{figure}
\plotone{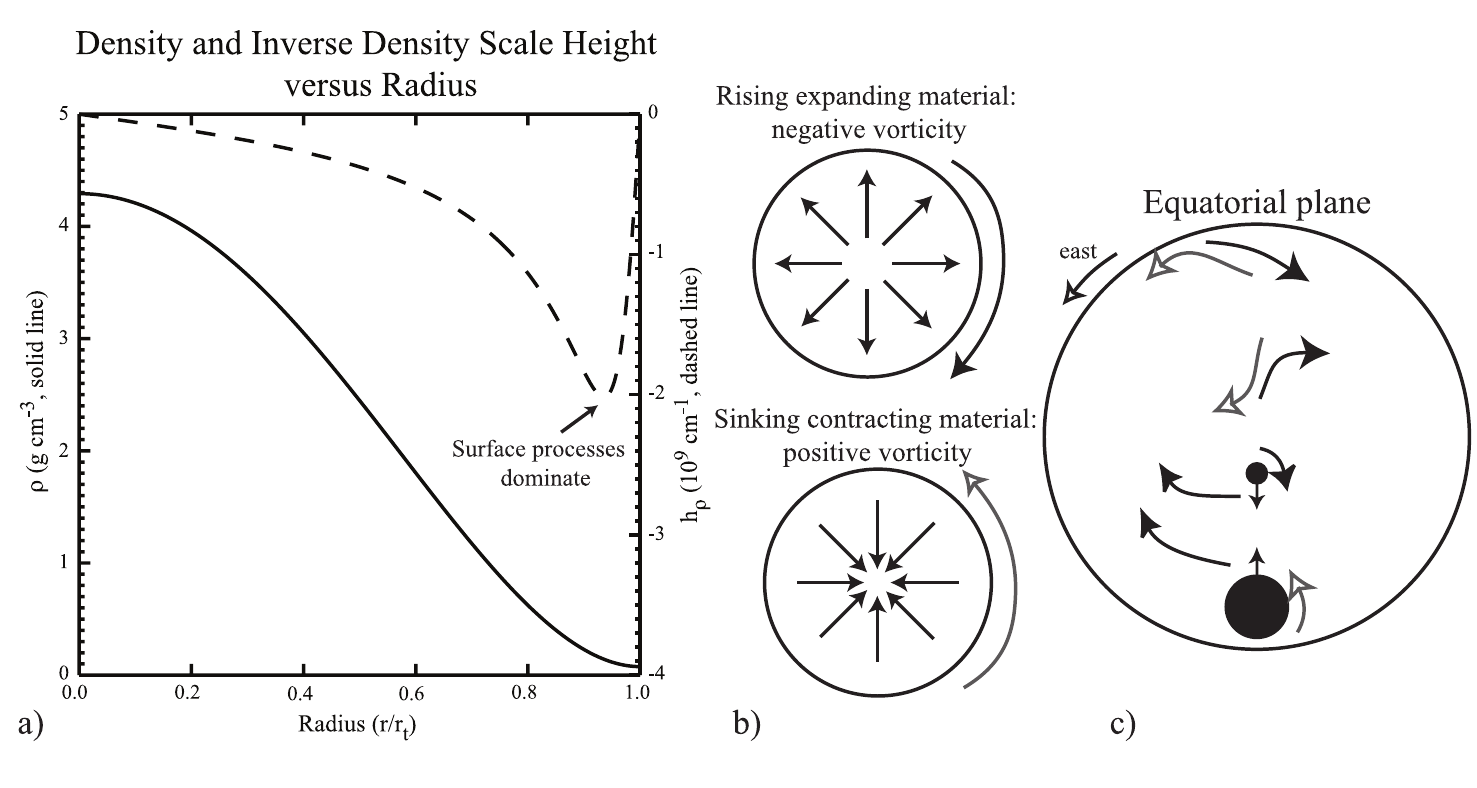}
\caption{Plot of a background density profile (solid line) and its corresponding inverse density scale heights, $h_{\rho}$ (dashed line), as a function of radius for a planet whose density drops off more quickly near the surface of its convective zone (a). Rising, expanding material generates negative vorticity due to Coriolis forces while sinking, contracting material generates positive vorticity (b). In (c) positive vorticity is advected towards both the inner and outer regions of the disk, however the dominant process occurs at the outer boundary due to that region's peak negative value of the inverse density scale height. Grey arrows indicate positive vorticity.}
\end{figure}

\begin{figure}
\plotone{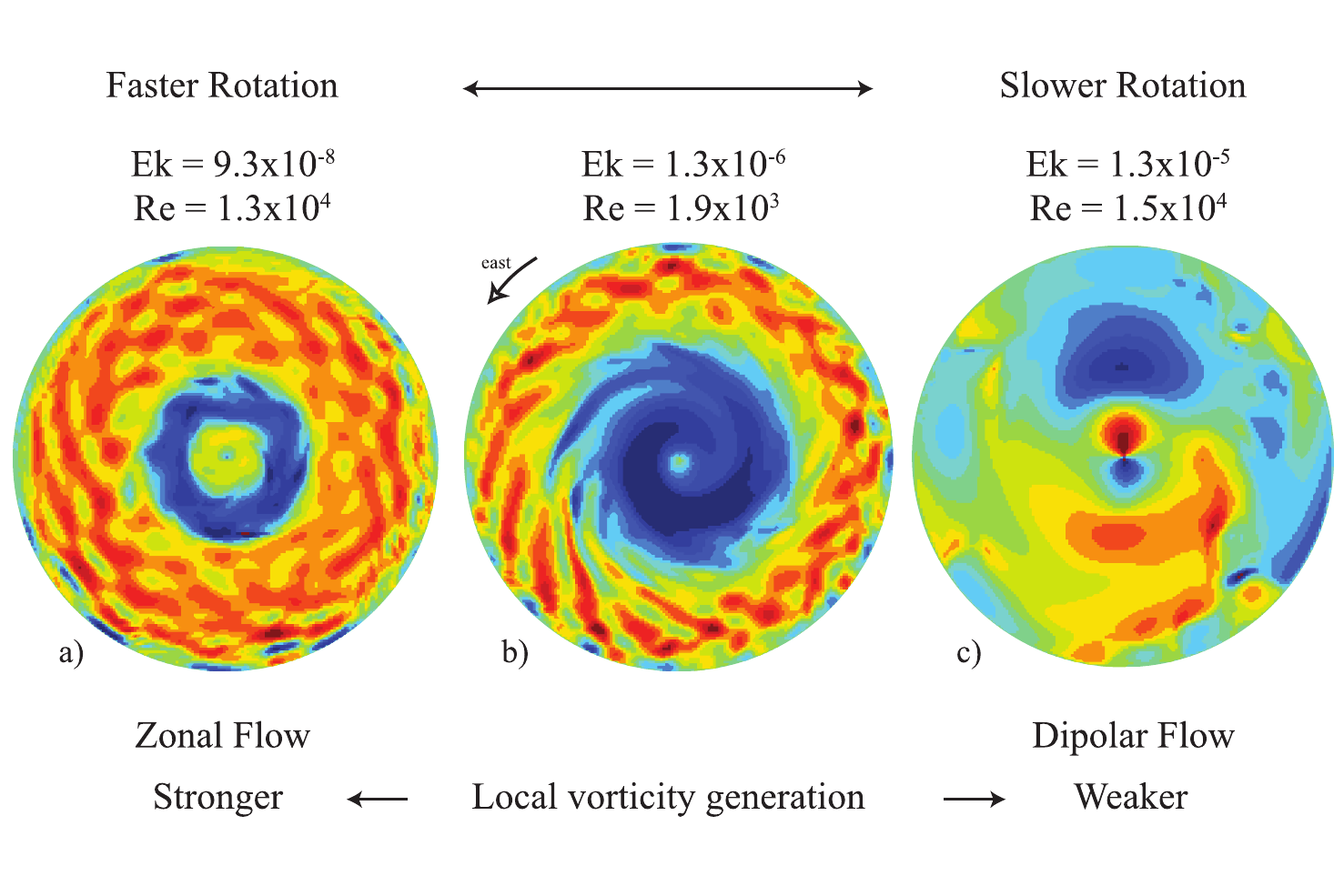}
\caption{Snapshots of the fluid flow in three representative cases. Red indicates prograde flow and blue retrograde flow; simulations are viewed from the north with prograde flow in the counterclockwise direction. The central case (b) shows a two-jet structure at an Ekman number of about $10^{-6}$. To the right, case (c) at an Ekman number $10$ times larger, the fluid has a dipolar flow pattern with flow directly through the center of the simulation. Conversely, to the left, case (a) is a simulation with an Ekman number $10$ times smaller than the central case, where three jets have formed.}
\end{figure}

\begin{figure}
\plotone{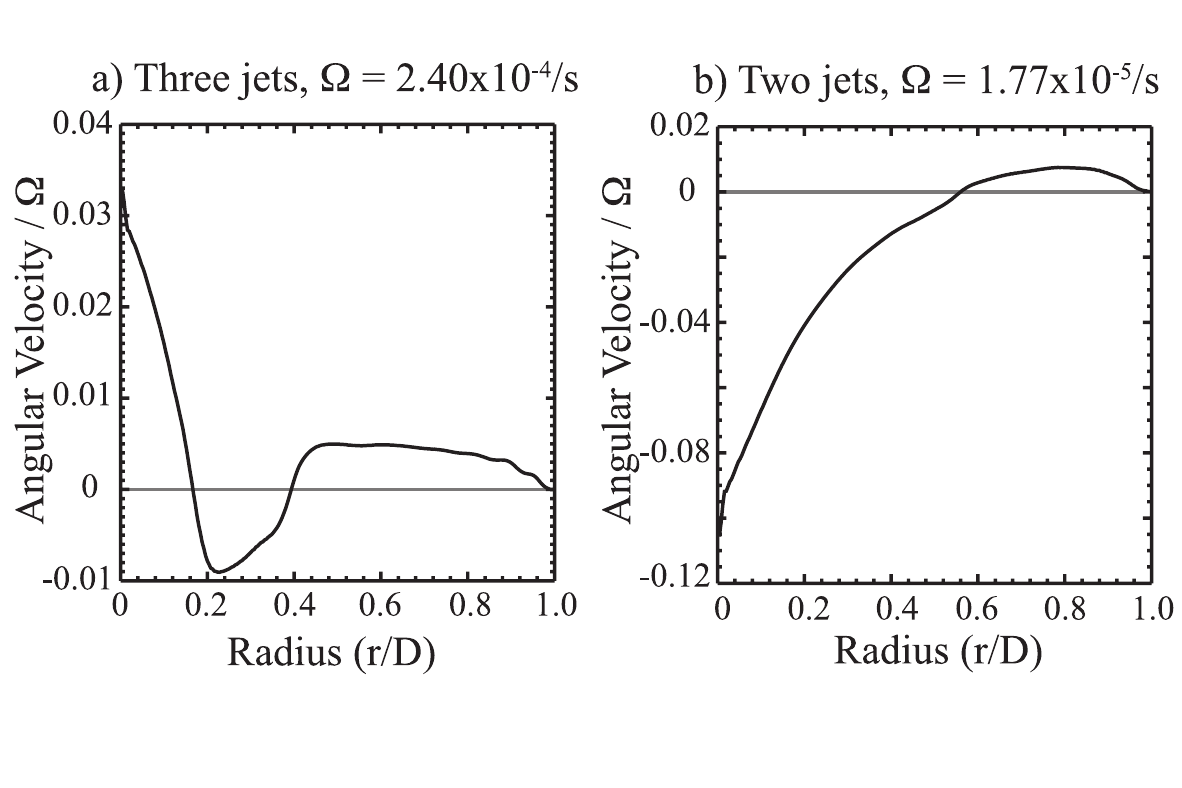}
\caption{The time and longitudinally-averaged angular velocity of two representative cases normalized to their respective rotation rates and plotted with respect to radius. Plot (a) corresponds to case (a) in Figure 2 with three jets, while plot (b) corresponds to case (b) in Figure 2 with two jets.}
\end{figure}

\begin{figure}
\plotone{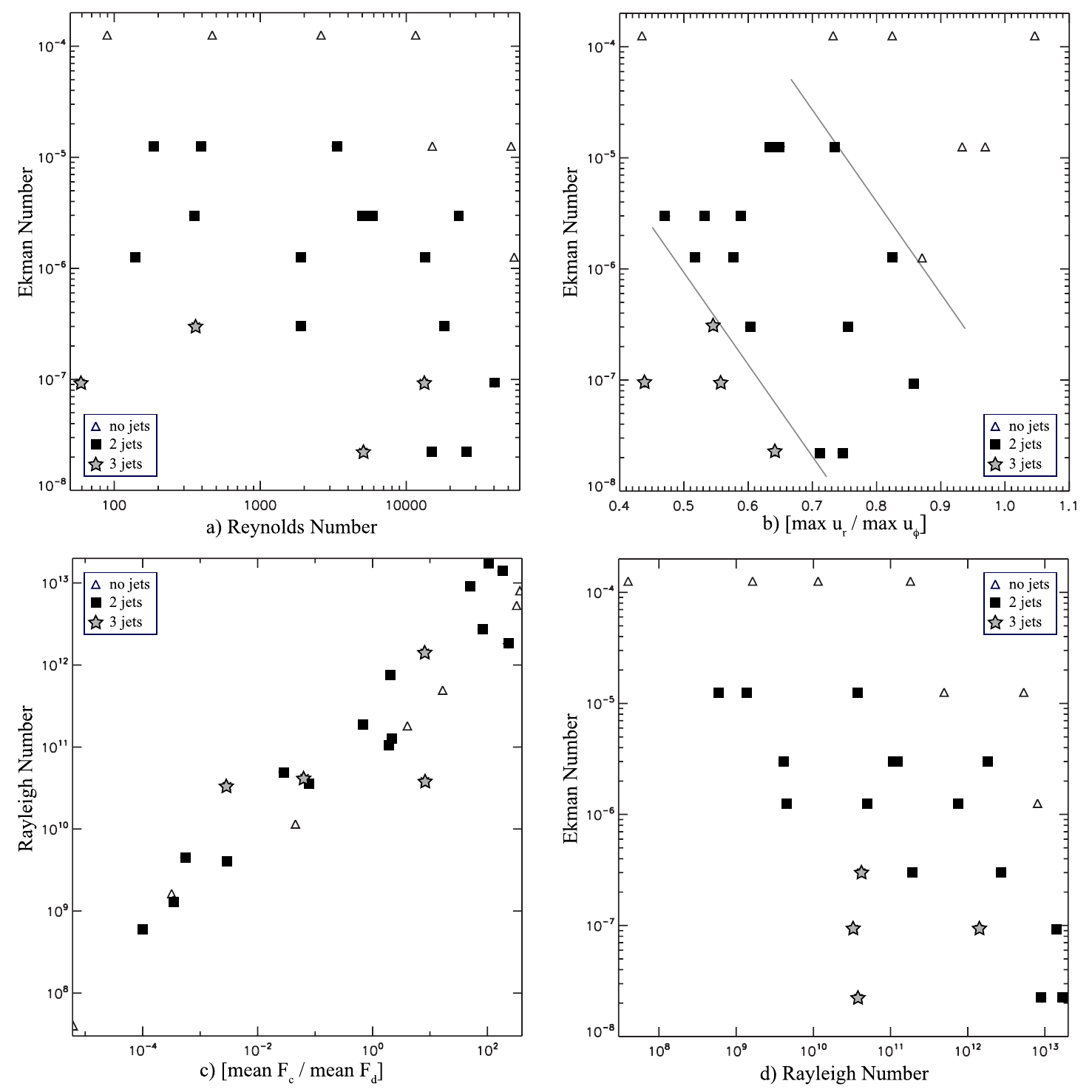}
\caption{Plot of Ekman number versus Reynolds number for the 26 simulations (a). Cases that did not form a zonal flow pattern are shown as open triangles, simulations with two jets as black squares, and simulations with three jets as stars. Plot of Ekman number versus the ratio of time-averaged maximum radial velocity to maximum tangential velocity (b). Lines have been added to show approximate divisions between the different groups. Plot of Rayleigh number versus the ratio of the mean convective heat flux to the mean diffusive heat flux (c). There is no apparent dependance on the number of jets present. Plot of Ekman number versus Rayleigh number (d). Increasing rotation rate leads to more jets while increasing Rayleigh number decreases the number of jets.}
\end{figure}

\begin{figure}
\plotone{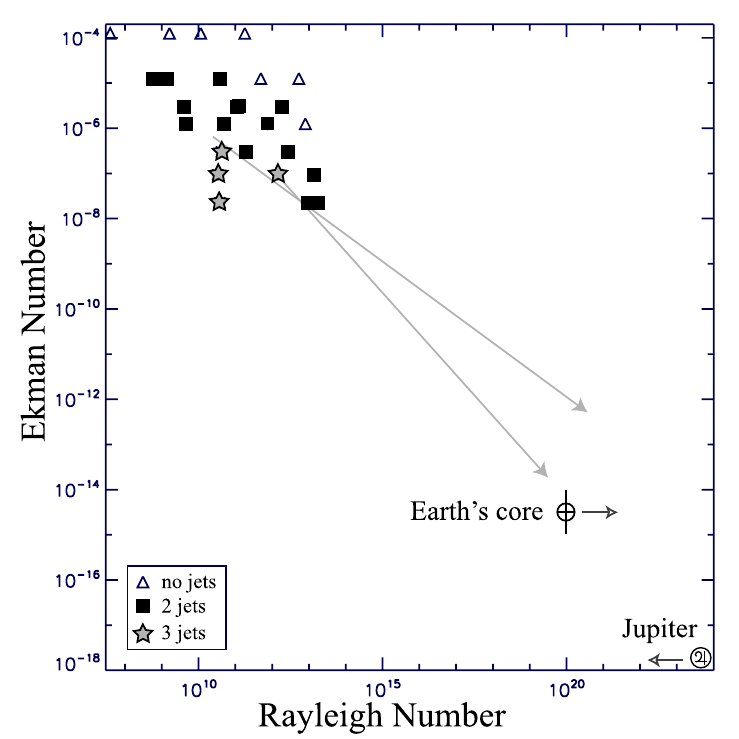}
\caption{Same as Figure 4d, but expanded to show the parameter space relevant to planetary interiors. Proposed parameter values for the EarthÕs liquid outer core and Jupiter are plotted for comparison. Simulations are still far, even in 2D, from the regimes in which these planets reside. Possible extrapolations for the transition between the two and three jet regions are shown in light grey. Simulations were done with four density scale heights, so the patterns seen should be shifted for the Earth (0.2 density scale heights) and Jupiter (over five density scale heights). Dark grey arrows show that the Earth would plot further to the right in a region where the local vorticity generation is less important and Jupiter would plot further to the left in a region where local vorticity generation is more important.}
\end{figure}



\end{document}